\documentclass[%
 reprint,
 superscriptaddress,
 amsmath,amssymb,
 aps,
]{revtex4-1}

\usepackage{graphicx}% Include figure files
\usepackage{dcolumn}% Align table columns on decimal point
\usepackage{multirow}
\usepackage{bm}% bold math
\usepackage{xcolor}
\usepackage{ulem}
\usepackage{mathrsfs}
\begin{document}

%\preprint{APS/123-QED}

\title{Low-$Q^{2}$ nucleon structure from an infrared-safe evolution scheme}
%% Force line breaks with \\
%%\thanks{A footnote to the article title}%

\author{Rong Wang}
\email{rwang@impcas.ac.cn}
\affiliation{Institute of Modern Physics, Chinese Academy of Sciences, Lanzhou 730000, China}
\affiliation{School of Nuclear Science and Technology, University of Chinese Academy of Sciences, Beijing 100049, China}


\date{\today}% It is always \today, today,

\begin{abstract}
The low-$Q^2$ nucleon structure is given with a simple nonperturbative input 
of three valence quarks and an all-order infrared-safe evolution scheme, 
showing some consistences with the experimental data. 
The resonance peaks in the experimental data of $F_2$ structure function are 
found to be modulated by the valence quark distributions. 
With little sea quark distribution at low $Q^2$
the valence-quark bump is clearly shown in the structure function. 
The three valence quark distributions at the hadronic scale is 
found to be the dominant origin of the PDFs at the hard scales.  
The high-twist corrections are needed to explain the sizeable discrepancy 
between the theory and the experimental measurements at low $Q^2$. 
The infrared-safe evolution scheme is a powerful tool 
for connecting the nucleon structures in the nonperturbative  
and perturbative regions. 
\end{abstract}

%% \pacs{21.60.Gx, 24.85.+p, 13.60.Hb}% PACS, the Physics and Astronomy
%% \keywords{Suggested keywords}%Use showkeys class option if keyword
\maketitle

%\tableofcontents

%%%%%%%%%%%%%%%%%%%%%%%%%%%%%%%%%%%%%%%%%%%%%%%%%%%%%%
\section{Introduction}\label{sec:intro}

Mapping the nucleon structure precisely is a hot topic in 
particle and nuclear physics, as it is closely related 
to the intricate and puzzling nonperturbative phenomena, 
such as the emergent hadron mass 
\cite{Ding:2022ows,Papavassiliou:2022wrb,Binosi:2022djx,Roberts:2021nhw,Carman:2023zke,Roberts:1994dr} 
and the color confinement 
\cite{Wilson:1974sk,Bander:1980mu,tHooft:1981bkw,Polyakov:1976fu,Mandelstam:1974pi}. 
With the advent of Electron-Ion Collider (EIC) 
in the near future \cite{AbdulKhalek:2021gbh,Accardi:2012qut}, 
the proposed Electron-ion collider in China (EicC) 
\cite{Chen:2018wyz,Chen:2020ijn,Anderle:2021wcy,Wang:2022xad} 
and JLab 24 GeV upgrade \cite{Accardi:2023chb}, the nucleon structure will 
be measured much more precisely in a broad kinematic region 
with multi-dimensional information, 
to finally tackle the challenging issues on the strong 
interaction in the nonperturbative region. 
The study of nucleon structure is essential for full understanding 
of the complex dynamics of the strong interaction, 
of which the underlying theory is quantum chromodynamics (QCD) 
\cite{Gross:1973id,Politzer:1973fx,Yang:1954ek,Fritzsch:1973pi,tHooft:1972tcz}. 

The core base for describing the nucleon structure is parton 
distribution function (PDF) $f(x)$ 
\cite{ParticleDataGroup:2022pth,Bjorken:1968dy,Feynman:1969ej,Bjorken:1969ja}, 
defined as the number density 
of the partons carrying $x$ fraction of the longitudinal momentum 
of the nucleon in the infinite-momentum frame (or light-cone momentum). 
Actually, the precision determinations of PDFs from global analyses 
\cite{Ball:2011uy,Pumplin:2002vw,Dulat:2015mca,Wang:2016sfq} 
have been a remarkable milestone of QCD theory. 
Yet, the origin of PDFs and the evolution of nucleon structure in 
the infrared region are not fully resolved. 
The nucleon structure can be easily and economically evaluated 
at low $Q^2$, for the degrees of freedom of sea quark and gluon 
fluctuations are frozen. However, how to link the calculations 
in the infrared region to the experimental measurements 
at high scales lacks comprehensive studies. 

There is no doubt that the nonperturbative effects should 
be implemented for the PDF evolution with the $Q^2$ scale 
in low-$Q^2$ domain. Recently, an infrared-safe evolution 
scheme is proposed to soundly extend the QCD evolution 
to $Q^2\approx 0$ GeV$^2$ \cite{Wang:2024wny}, with the effects of effective parton 
mass, saturating running strong coupling, and parton-parton recombination. 
This kind of evolution is also argued to be all-order exact \cite{Yin:2023dbw}. 
Under this novel evolution scheme, the PDFs are delivered 
consistently in the whole $Q^2$ range from 0 GeV$^2$ to infinity. 
With the help of the new evolution scheme, firstly, we would like 
to check the origin of PDFs. According to the quark model assumption, 
the nucleon is made of three quarks, which are the minimum number 
of quarks forming a color singlet. Without the contributions of 
sea quark and gluon distributions, the bump of valence quark distribution 
should shown up in the structure function at extremely low $Q^2$. 
Secondly, for example, the valence quark distributions estimated 
in infrared region from maximum entropy method \cite{Wang:2014lua} are tested 
using the suggested infrared-safe evolution equations. 
The third aim of this study is to check the violation of 
factorization \cite{Collins:1987pm,Collins:1989gx,Sterman:1995fz},  
via the comparisons between the theoretical predictions  
and the experimental measurements of structure function $F_2$ 
at the low $Q^2$ scales.

The organization of the paper is as follows. 
The input valence quark distributions at the hadronic scale 
and the nonperturbative effects within the Continuum 
Schwinger Methods (CSMs) \cite{Roberts:1994dr} are discussed in Sec. \ref{sec:nonperturbative}. 
How the nonperturbative effects modify the evolution equations 
and the elicited infrared-safe evolution equations are demonstrated 
in Sec. \ref{sec:infrared-safe-evolution}. 
The predicted $F_2$ structure functions at low $Q^2$ are shown 
and compared with JLab data \cite{Armstrong:2001xj} in Sec. \ref{sec:results}. 
The discussions on the origin of PDFs and the factorization 
invalidity at low $Q^2$ are also provided in Sec. \ref{sec:results}. 
A concise summary is provided in Sec. \ref{sec:summary}.

%%%%%%%%%%%%%%%%%%%%%%%%%%%%%%%%%%%%%%%%%%%%%%%%%%%%%%
\section{Nonperturbative effects}\label{sec:nonperturbative}

All the complex strong interaction dynamics at long distance are 
reflected in the nonperturbative input of PDFs. 
The nonperturbative input is the initial PDFs at the hadronic scale, 
and the hadronic scale is defined as the scale where the nucleon 
contains only the valence and minimal components. 
The nonperturbative input is related to the nucleon wave function 
at the low energy scale governed by the nonperturbative strong interaction. 

Dynamical chiral Symmetry Breaking (DCSB) is a prominent phenomenon 
in QCD \cite{Roberts:2021nhw,Lane:1974he,Politzer:1976tv,Cornwall:1989gv,
Nambu:1961tp,Krein:1990sf,Chang:2011zgy,Roberts:2007ji,Munczek:1994zz}, 
which offers a mechanism of generating mass from nothing. 
In the chiral limit, a quark acquires the massive mass through  
the strong interactions with gluons at the low momentum scale. 
By evaluating the two-point Green function of quark, 
both CSMs and LQCD give the excellent and mutual consistence 
regarding the features of DCSB \cite{Chang:2011zgy}. 
The large effective mass of quark definitely affects 
the splitting kernels of the evolution equations 
that are derived under the assumption of massless quarks. 
DCSB is one unavoidable nonperturbative effect in the infrared region. 

To perform the calculations in the infrared region, 
an infrared finite strong coupling is needed. 
According to the recent definition of QCD effective charge 
\cite{Binosi:2016nme,Cui:2019dwv}, 
the running strong coupling saturates up to a finite value 
in the infrared region, for the Landau pole is tamed  
due to the massive effective mass of gluon at low momentum. 
The saturating $\alpha_{\rm s}$ is one important ingredient 
of our calculations.

\subsection{Nonperturbative input of PDFs} 

\begin{figure}[htbp]
\begin{center}
\includegraphics[width=0.43\textwidth]{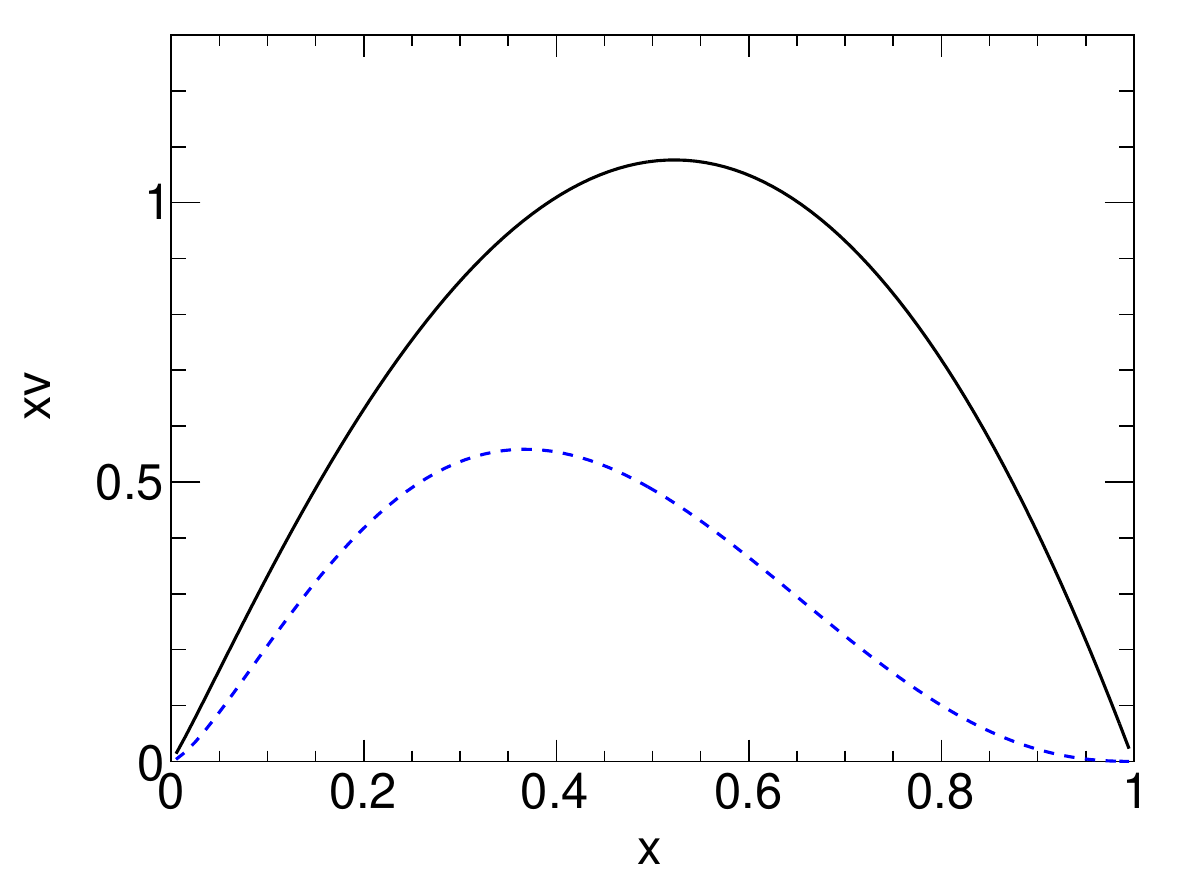}
\caption{The input parton distributions of mere valence quarks 
at the sufficiently low $Q_0^2$ scale. The solid and dashed curves 
show the up and down valence quark distributions respectively. }
\label{fig:iput_valence_distributions}
\end{center}
\end{figure}

It is very convenient to compute the nucleon structure at an extremely 
low $Q^2$ scale by reason of the minimal constituents 
inside the nucleon at the scale. 
Nowadays, the nonperturbative input can be estimated with the 
first principle techniques of CSMs \cite{Yu:2024qsd,Lu:2022cjx,Chang:2014lva,Ding:2019lwe,Cui:2020tdf} 
and Lattice QCD simulation 
\cite{Ji:2013dva,Ji:2014gla,Ji:2020ect,Ji:2017oey,Ma:2014jla,Ma:2017pxb},  
and many other model calculations 
\cite{Wang:2014lua,Lan:2019rba,Mondal:2019jdg,
Holt:2010vj,Mineo:2003vc,Kock:2020frx,Steffens:1994hf,Radyushkin:2004mt}. 
For the simplicity of calculation, the nonperturbative input 
from the maximum entropy method (MEM) \cite{Wang:2014lua} is used in this work. 
Actually, there is no much difference by using different nonperturbative inputs. 

Fig. \ref{fig:iput_valence_distributions} shows the simplest nonperturbative 
input of just three valence quarks at extremely low $Q^2$ scale, 
where only the valence components in the nucleon can be resolved or probed. 
The input three valence quark distributions is evaluated with MEM, 
under the constraints of the quark model assumption 
and the confinement radius of quarks \cite{Wang:2014lua}. 
This MEM nonperturbative input is found to be more or less 
consistent with the global fits and the experimental data \cite{Wang:2014lua}.

\subsection{Dressed parton mass} 

\begin{figure}[htbp]
\begin{center}
\includegraphics[width=0.43\textwidth]{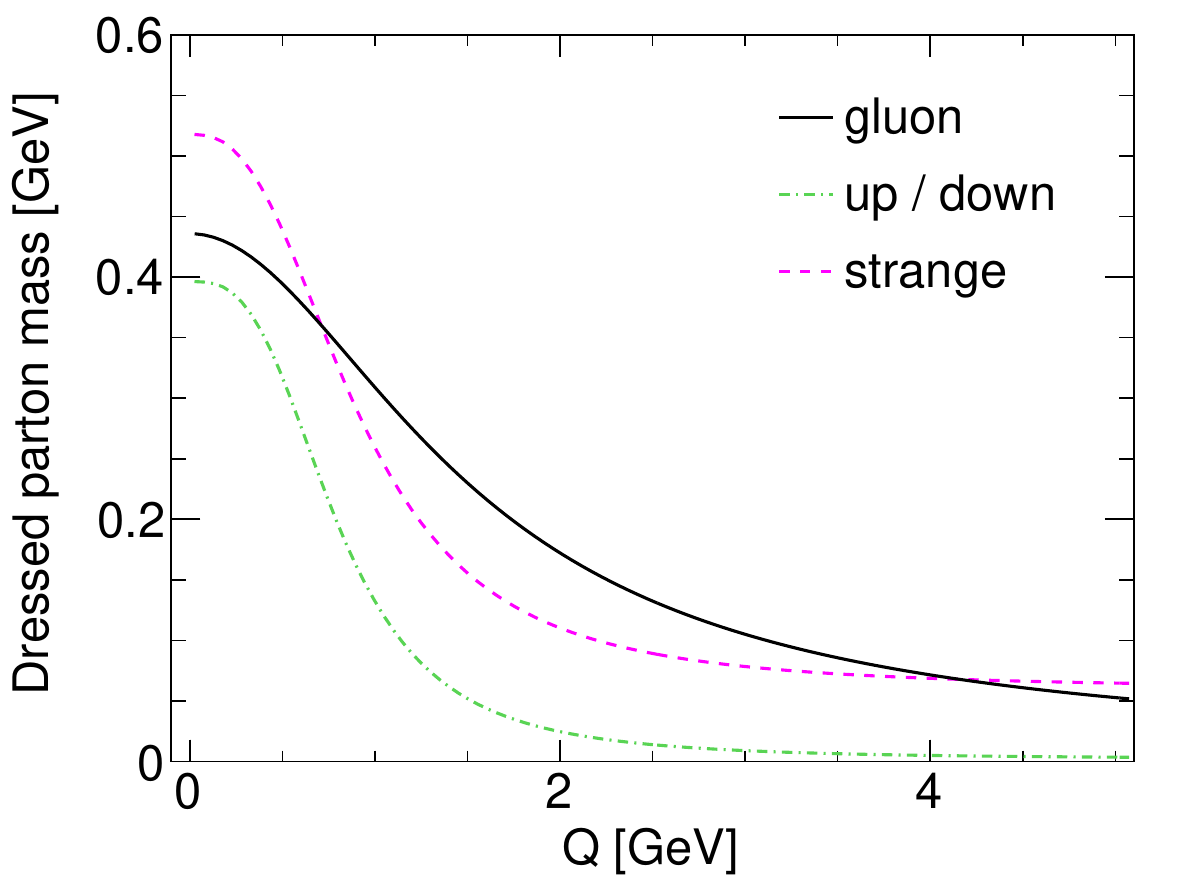}
\caption{The dressed parton mass as a function of the momentum \cite{Roberts:2021nhw,Aguilar:2019uob}. 
The types of the partons are indicated in the figure.  }
\label{fig:dressed_parton_mass}
\end{center}
\end{figure}

The effective masses of quark and gluon are important parameters for 
the modifications of splitting kernels of QCD evolution equations 
in the infrared region. For DCSB, the quark mass is a smooth function 
of the momentum. The mass functions of quark and gluon can be 
evaluated with Dyson-Schwinger equations which take into account 
the contributions of infinite diagrams \cite{Roberts:2021nhw,Aguilar:2019uob}. 
Usually, the symmetry-preserving 
truncation of the diagrams (Rainbow-Ladder truncation) is applied 
in the calculations \cite{Munczek:1994zz,Bender:1996bb}. 
Nevertheless, Dyson-Schwinger equations 
predict the quark mass function that is amazingly consistent with LQCD simulations. 

Fig. \ref{fig:dressed_parton_mass} shows the dressed quark 
and gluon mass functions from CSMs \cite{Roberts:2021nhw,Aguilar:2019uob}. 
One sees that the dressed quark mass decreases 
monotonically with the increasing momentum. 
The constituent quark mass is smoothly connected 
with the current quark mass under the DCSB mechanism 
of complex strong interaction dynamics. 
In this work, the parametrization forms of these mass functions 
are used in the following calculations of 
the infrared-safe evolution equations.

\subsection{Saturating running strong coupling} 

\begin{figure}[htbp]
\begin{center}
\includegraphics[width=0.43\textwidth]{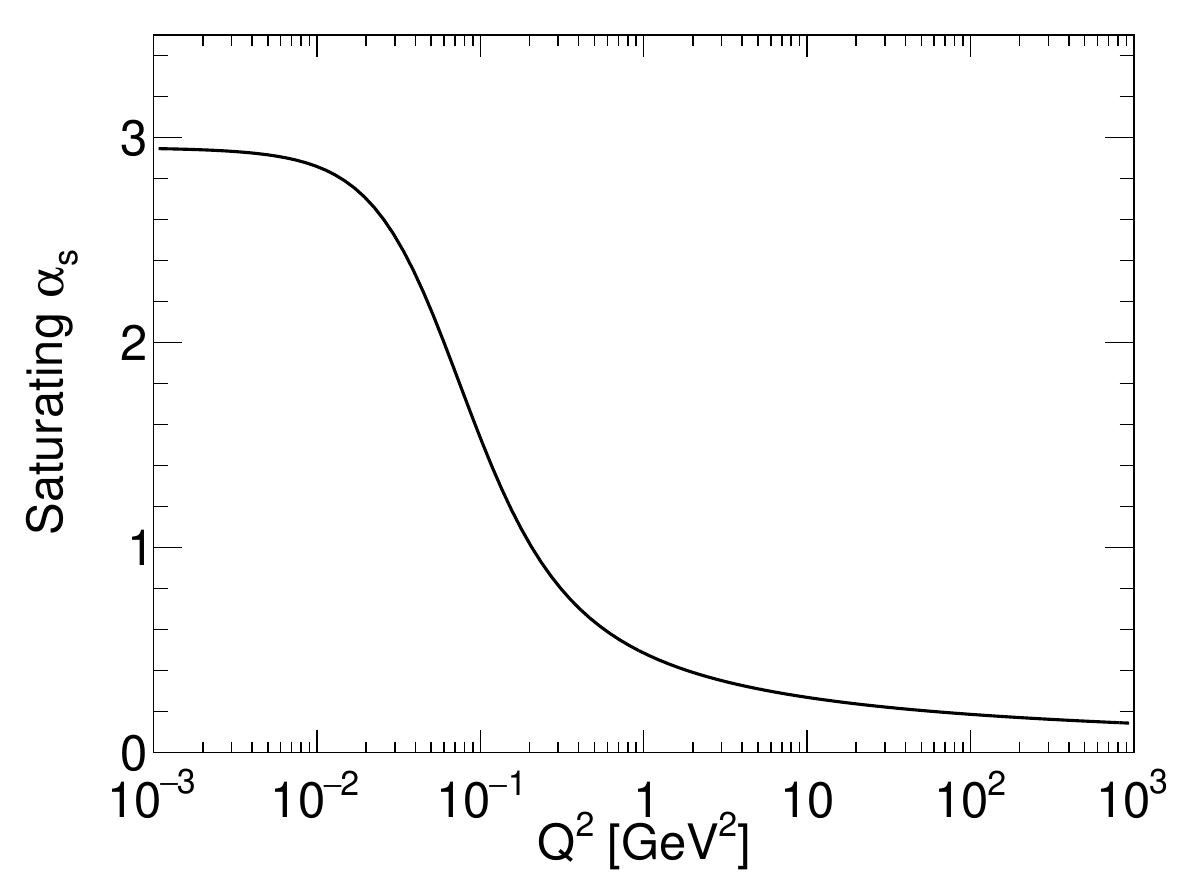}
\caption{The saturating running strong coupling $\alpha_{\rm s}$ 
as a function of the $Q^2$ scale \cite{Roberts:2021nhw,Cui:2020tdf,Binosi:2016nme,Cui:2019dwv}. 
The running coupling saturates in the far-infrared region around $\pi$.  }
\label{fig:saturating_alphas}
\end{center}
\end{figure}

The running strong coupling $\alpha_{\rm s}$ is a key parameter 
for any QCD processes. To get rid of the divergent problem 
of perturbative $\alpha_{\rm s}$, we capitalize a 
renormalization-group-invariant and process-independent effective 
charge in QCD \cite{Roberts:2021nhw,Cui:2020tdf,Binosi:2016nme,Cui:2019dwv}.  
Based on the concept of QCD effective charge, 
the infrared behavior of $\alpha_{\rm s}$ is quite different 
form the traditional perturbative $\alpha_{\rm s}$. 
The QCD effective charge derives a saturating $\alpha_{\rm s}$ 
which is free of Landau pole. 
The screening effect is suppressed at long distance due 
to the big effective mass of the gluon. 
In this work, the saturating $\alpha_{\rm s}$ is applied 
in the calculations of infrared-safe evolution equations. 

Fig. \ref{fig:saturating_alphas} shows the saturating $\alpha_{\rm s}$ from CSMs 
\cite{Roberts:2021nhw,Cui:2020tdf,Binosi:2016nme,Cui:2019dwv}. 
This saturating $\alpha_{\rm s}$ provides an infrared completion 
with $\alpha_{\rm s}/\pi\sim 0.97$ in the infrared limit. 
At high $Q^2$, the saturating $\alpha_{\rm s}$ agrees well 
with the perturbative $\alpha_{\rm s}$. 
According to the recent calculations of CSMs, 
a parametrization of saturating $\alpha_{\rm s}$ is given by
\cite{Binosi:2016nme,Cui:2019dwv,Cui:2020tdf,Roberts:2021nhw},
\begin{equation}
\begin{split}
   \alpha_{\rm s} = \frac{\gamma_m\pi}{{\rm ln}[\mathscr{K}^2(k^2)/\Lambda_{\rm QCD}^2]}, \\
   \mathscr{K}(y) = \frac{a_0^2 + a_1y + y^2}{b_0+y}, \\
\end{split}
\label{eq:effective-charge}
\end{equation}
where the parameters are $\gamma_m=4/\beta_0$, $\beta_0=11-(2/3)n_f$, 
and $\Lambda_{\rm QCD}=0.234$ GeV.

%%%%%%%%%%%%%%%%%%%%%%%%%%%%%%%%%%%%%%%%%%%%%%%%%%%%%%
\section{An infrared-safe evolution scheme}\label{sec:infrared-safe-evolution}
%%%%%%%%%%%%%%%%%%%%%%%%%%%%%%%%%%%%%%%%%%%%%%%%%%%%%%

Considering the nonperturbative effects discussed above, 
we have proposed a novel and infrared-safe evolution scheme \cite{Wang:2024wny}. 
The massive parton mass hinders the parton radiation processes 
in Dokshitzer-Gribov-Lipatov-Altarelli-Parisi (DGLAP) 
evolution \cite{Dokshitzer:1977sg,Gribov:1972ri,Altarelli:1977zs} 
over the $Q^2$ scale. 
The evolution speed is controlled with a factor $Q^2/(Q^2+M_q^2)$, 
where $M_q$ is the quark effective mass provided by DCSB. 
The saturating $\alpha_{\rm s}$ offers us a finite interaction 
strength in QCD at any given momentum scale. 
The last ingredient for the infrared-safe evolution scheme 
is the parton-parton recombination correction 
\cite{Gribov:1983ivg,Mueller:1985wy,Zhu:1998hg}.  
The parton-parton recombination process balances the parton splitting process, 
slowing down the growth of gluon distribution at small $x$, 
ensuring the cross section unitarity restored at high energy. 
Due to the large parton size at low $Q^2$, the parton-parton 
overlapping effect should not be ignored at the low scales.

With the implementations of the nonperturbative effects and 
the parton-parton overlapping effect, 
we finally get an infrared-safe evolution scheme which 
works at any given $Q^2$ scale. 
In this work, the infrared-safe evolution scheme is applied 
to a MEM nonperturbative input, so as to deliver the nucleon 
PDFs at the low $Q^2$ scales. 
The infrared-safe evolution equations are written as,  
\begin{equation}
\begin{split}
   \frac{dxq_{\rm i}^{\rm NS}}{d{\rm ln}(Q^2)} =
   \frac{Q^2}{Q^2+M_{\rm q}^2}\frac{\alpha_{\rm s}(Q^2)}{2\pi}
   P_{\rm qq} \otimes xq_{\rm i}^{\rm NS},
\end{split}
\label{eq:modifiedAP-NS}
\end{equation}
for the flavor non-singlet quark distributions, 
\begin{equation}
\begin{split}
   \frac{dx\bar{q}_{\rm i}}{d{\rm ln}(Q^2)} =
   \frac{Q^2}{Q^2+M_{\rm q}^2}\frac{\alpha_{\rm s}(Q^2)}{2\pi}
   \left[
   P_{\rm qq} \otimes x\bar{q}_{\rm i} + P_{\rm qg} \otimes xg
   \right] \\
   -\frac{Q^2}{Q^2+M_{\rm q}^2}\frac{\alpha_{\rm s}^2(Q^2)}{4\pi R^2Q^2}\int_{x}^{1/2}
   \frac{dy}{y}xP_{\rm gg\rightarrow\bar{q}}(x,y)[yg(y,Q^2)]^2 \\
   +\frac{Q^2}{Q^2+M_{\rm q}^2}\frac{\alpha_{\rm s}^2(Q^2)}{4\pi R^2Q^2}\int_{x/2}^{x}
   \frac{dy}{y}xP_{\rm gg\rightarrow\bar{q}}(x,y)[yg(y,Q^2)]^2,
\end{split}
\label{eq:modifiedAP-sea}
\end{equation}
for the sea quark distributions, and 
\begin{equation}
\begin{split}
   \frac{dxg}{d{\rm ln}(Q^2)} =
   \frac{Q^2}{Q^2+M_{\rm g}^2}\frac{\alpha_{\rm s}(Q^2)}{2\pi}
   \left[
   P_{\rm gq} \otimes x\Sigma + P_{\rm gg} \otimes xg
   \right] \\
   -\frac{Q^2}{Q^2+M_{\rm g}^2}\frac{\alpha_{\rm s}^2(Q^2)}{4\pi R^2Q^2}\int_{x}^{1/2}
   \frac{dy}{y}xP_{\rm gg\rightarrow g}(x,y)[yg(y,Q^2)]^2 \\
   +\frac{Q^2}{Q^2+M_{\rm g}^2}\frac{\alpha_{\rm s}^2(Q^2)}{4\pi R^2Q^2}\int_{x/2}^{x}
   \frac{dy}{y}xP_{\rm gg\rightarrow g}(x,y)[yg(y,Q^2)]^2,
\end{split}
\label{eq:modifiedAP-g}
\end{equation}
for the gluon distribution, 
where $P_{\rm qq}$, $P_{\rm qg}$, $P_{\rm gg}$, $P_{\rm gq}$ are 
the standard splitting kernels, and $P_{\rm gg\rightarrow\bar{q}}$, 
$P_{\rm gg\rightarrow g}$ are the gluon-gluon recombination kernels. 
The factor $1/(4\pi R^2)$ in the equations is for two-parton density normalization, 
and $R$ is the correlation length of two overlapping partons 
fixed already in a previous study \cite{Wang:2016sfq}.

In above equations, the factor $Q^2/(Q^2+M_{\rm q/g}^2)$ quenches the evolution 
at low $Q^2$, where $M_{\rm q/g}(Q^2)$ denote the mass functions offered from CSMs. 
The factor $Q^2/(Q^2+M_{\rm q/g}^2)$ comes from the modifications of 
the standard evolution kernels (the splitting and recombination processes). 
As $Q^2/(Q^2+M_{\rm q/g}^2)$ is not $x$-dependent, 
we take it out from the integrals on the right sides of the equations. 
The $\alpha_{\rm s}(Q^2)$ is the saturating $\alpha_{\rm s}$
given in Eq. (\ref{eq:effective-charge}). It is finite at any $Q^2$ scale. 
The terms with $\int_x^{1/2}$ come from the gluon-gluon recombination 
processes which reduce the growth of parton density. 
The terms with $\int_{x/2}^{x}$ are the anti-shadowing contributions 
to keep the momentum conservation in the gluon-gluon recombination processes.

%%%%%%%%%%%%%%%%%%%%%%%%%%%%%%%%%%%%%%%%%%%%%%%%%%%%%%
\section{Results and discussions}\label{sec:results}
%%%%%%%%%%%%%%%%%%%%%%%%%%%%%%%%%%%%%%%%%%%%%%%%%%%%%%

\begin{figure}[htbp]
\begin{center}
\includegraphics[width=0.43\textwidth]{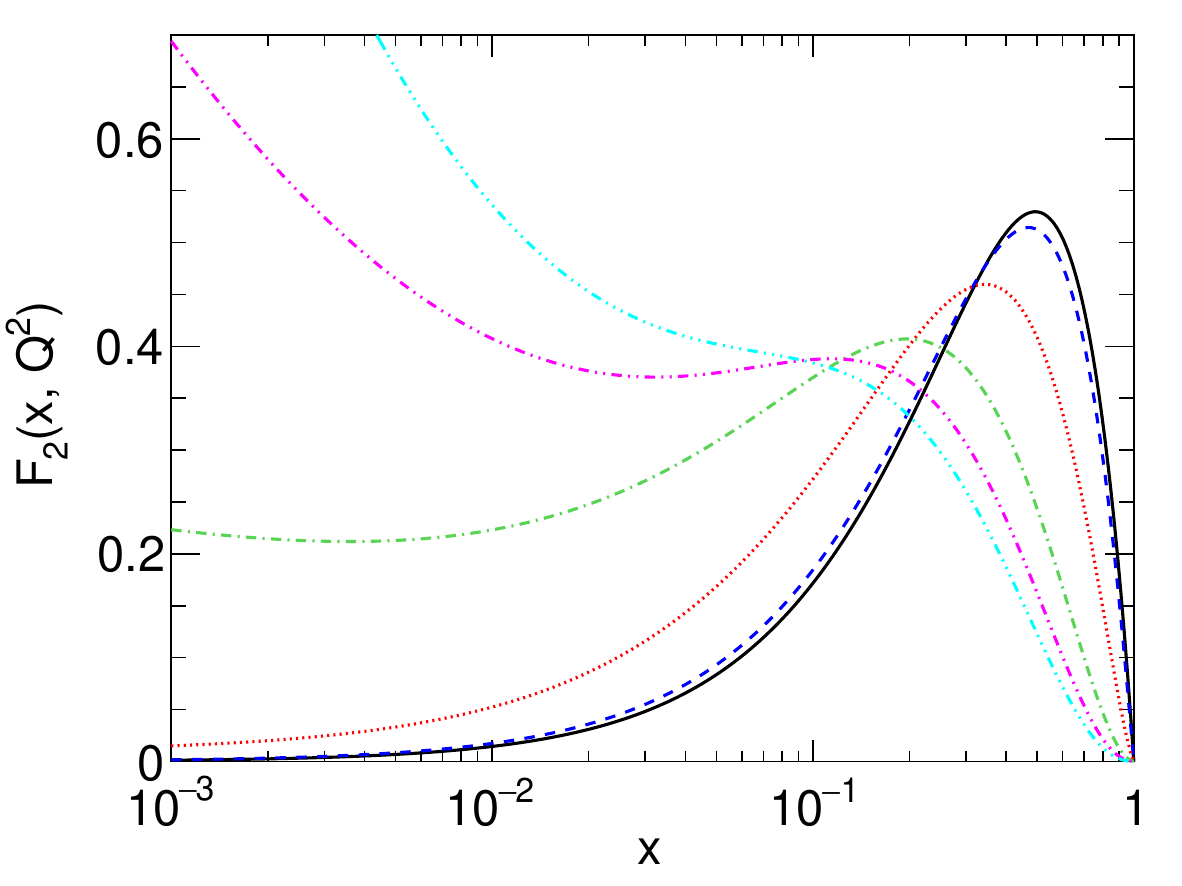}
\caption{The obtained $F_2$ structure function as a function of $x$ 
at different scales from the infrared-safe evolution equation. 
The $Q^2$ scales are at 0.001 (solid), 0.01 (dashed), 0.1 (dotted), 
1 (dashed-dotted), 10 (dashed-dotted-dotted), 100 GeV$^2$ 
(dashed-dotted-dotted-dotted), respectively. }
\label{fig:F2_evolution}
\end{center}
\end{figure}

One prominent feature of the infrared-safe evolution scheme is 
the freeze of PDF evolution as $Q^2$ approaching zero, 
explicitly indicated with the factor $Q^2/(Q^2+M_{\rm q/g}^2)$. 
It is found that the evolved PDFs are not sensitive 
to the choice of input hadronic scale $Q_0^2$, 
so long as $Q_0^2 \lesssim 0.1$ GeV$^2$. 
Owing to the freeze of PDF evolution in the region of 
$Q^2\approx 0$ GeV$^2$, a naive and natural choice of 
the hadronic scale is $Q_0^2= 0$ GeV$^2$. 
In the following calculations, we set $Q_0^2= 0.0001$ GeV$^2$ 
for a fast calculation of PDFs to a high $Q^2$ scale.

The infrared-safe evolution equation is applied to the 
three valence quark distributions from MEM at $Q_0^2= 0.0001$ GeV$^2$, 
to obtain the PDFs at higher $Q^2$. 
With the evolved PDFs, the proton structure functions are then computed 
at different $Q^2$ scales, which are shown in Fig. \ref{fig:F2_evolution}. 
One sees that the structure function peaks in the valence region at low 
$Q^2$ scales, due to the tiny sea quark distribution at low scale. 
Thus, the valence-quark bump in $F_2$ structure function at low $Q^2$ is 
a strong proof of that the dominant valence quark distributions are 
the origin of the PDFs measured at the hard scales. 
In other words, the quark model description of the nucleon is the origin 
of the asymptotic partons in the infinite momentum frame.

\begin{figure*}[htbp]
\begin{center}
\includegraphics[width=0.8\textwidth]{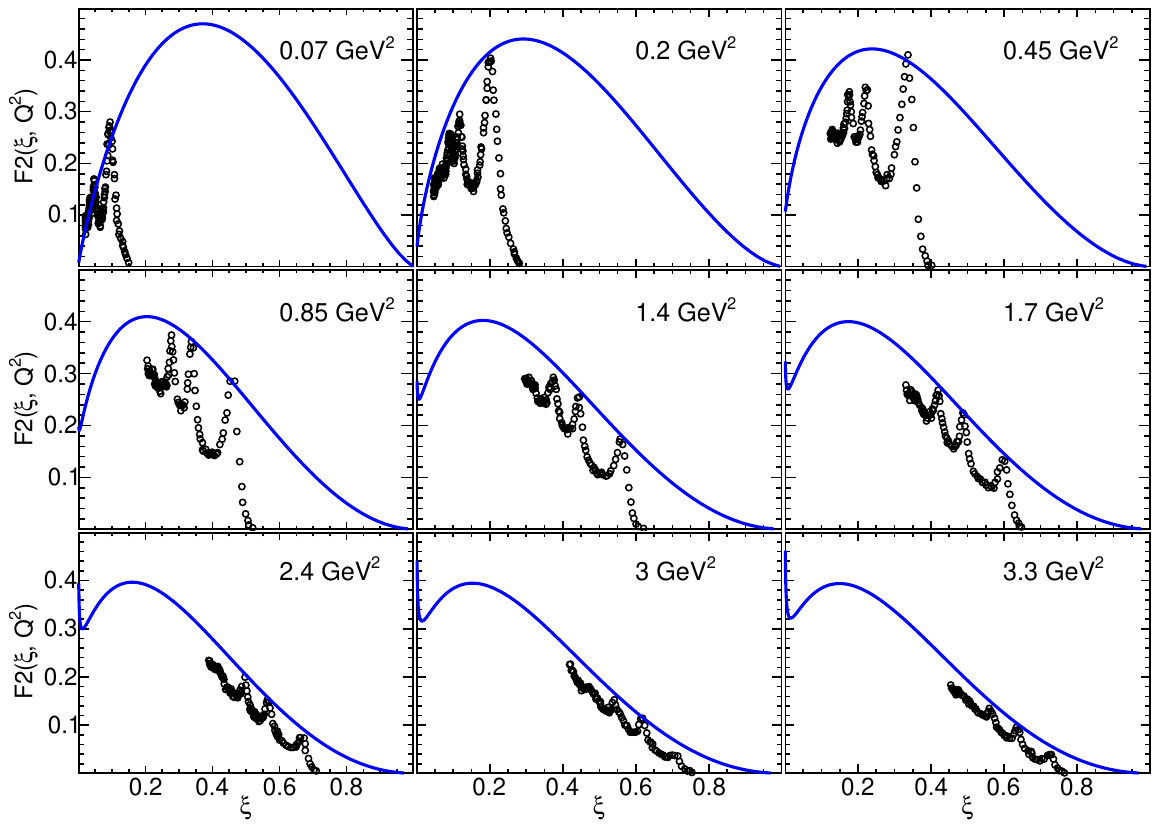}
\caption{The $F_2$ structure function as a function of Nachtmann variable 
$\xi$ at low $Q^2$ scales in the large $x$ region. 
The circles show the experimental data from JLab \cite{Armstrong:2001xj}. 
The curves show the predictions from the infrared-safe evolution scheme.  }
\label{fig:F2_vs_xi_at_lowQ2}
\end{center}
\end{figure*}

Our predicted structure functions are compared to JLab measurement \cite{Armstrong:2001xj}
at the low $Q^2$ scales in the large $x$ region, 
which is shown in Fig. \ref{fig:F2_vs_xi_at_lowQ2}. 
At finite $Q^2$, the target mass effect modifies the identification of the 
Bjorken variable $x$ with the light-cone momentum fraction. 
It was argued that the effect of the finite target mass can be removed 
by analyzing the structure functions in terms of the Nachtmann variable 
\cite{Nachtmann:1973mr,Schienbein:2007gr},  
which is defined as $\xi=2x/[1+\sqrt{1+4x^2M_{\rm N}^2/Q^2}]$, 
where $M_{\rm N}$ is the nucleon mass. 
One sees that at high $Q^2$, $\xi$ and $x$ are almost identical. 
Therefore, we show the measured structure function at low $Q^2$ as a function of $\xi$. 
In Fig. \ref{fig:F2_vs_xi_at_lowQ2}, one finds that the experimental data 
is obviously lower than the theoretical predictions. 
However, the resonance peaks in the experimental data are amazingly 
well modulated with our theoretical predictions. 
This indicates that the quark-hadron duality exists at low $Q^2$ 
and our theoretical calculations are consistent with 
the experimental data approximately.  

Judged from the JLab data at different $Q^2$ \cite{Armstrong:2001xj}, 
one sees clearly the valence-quark bump in the measured structure function, 
regardless the fluctuations caused by the resonances. 
The heights of the resonance peaks are well described with 
the overwhelming valence quark distribution in the low-$Q^2$ region. 
According to quark-hadron duality, the experimental data at low $Q^2$ 
can be interpreted roughly with the three valence quark distributions. 
To conclude, the quark model description in the infrared region is 
the dominant origin for the asymptotic partons observed 
at high energies.

In fact, we could modify the input three valence quark distributions 
to explain the low-$Q^2$ data slightly better. 
Nonetheless, the inherent inconsistency can not be removed by 
just tuning the nonperturbative input at $Q_0^2$. 
Moreover, the lower the $Q^2$ is, the larger the discrepancy 
between the theory and the data is found. 
The theoretical calculations are provided under the factorization theorem. 
Thereby, from this study, we conclude that the lower the $Q^2$ is,  
the more significant violation of the factorization theorem is found. 
Of course, this finding is not surprising, 
since at low $Q^2$ the sizeable high-twist correction is 
not included in the factorization formula. 
To improve the interpretations of the low-$Q^2$ data, 
the study of high-twist contributions is quite necessary.

%%%%%%%%%%%%%%%%%%%%%%%%%%%%%%%%%%%%%%%%%%%%%%%%%%%%%%%%%%%%%%%%%%%%%%%%%%%%%%%%
\section{Summary}\label{sec:summary}
%%%%%%%%%%%%%%%%%%%%%%%%%%%%%%%%%%%%%%%%%%%%%%%%%%%%%%%%%%%%%%%%%%%%%%%%%%%%%%%%

An infrared-safe evolution scheme is applied to the 
initial three valence quark distributions at $Q_0^2$, 
to study the origin of PDFs. 
The three valence quarks in the quark model is found 
to be the dominant origin of the exquisite PDFs at the hard scales. 
Regardless of the structures of resonances in the data, 
the JLab experiment has shown a clear valence-distribution bump 
in the measured $F_2$ structure function at low $Q^2$ scales. 
The quark-hadron duality is observed as well.

The infrared-safe evolution scheme is an essential bridge 
for connecting the nucleon structures in the nonperturbative 
region and perturbative region. 
The massive dressed quark mass, the saturating $\alpha_{\rm s}$, 
and the parton-parton overlapping corrections are important 
for the validity and success of the infrared-safe evolution scheme. 
Via the infrared-safe evolution scheme, 
the quark model is smoothly linked to the parton model, 
and the nucleon PDFs can be delivered in the whole momentum range 
from 0 GeV$^2$ to any given high scales. 

In this study, the factorization violation at low $Q^2$ is observed 
and tested with the JLab experimental data. 
Consistent with the suppression behavior of high-twist contributions, 
the higher the $Q^2$ is, the smaller the violation is seen. 
The quantitative study of the high-twist phenomenon is looking forward 
to improve the theoretical explanation of the low-$Q^2$ data. 
Although the factorization is violated at low $Q^2$, 
the PDFs still exist at low $Q^2$ guaranteed by the definition 
of the matrix elements of bilocal light-cone operators. 
By extracting the leading-twist contribution to the structure function 
from the PDFs, we could get some valuable information on 
the high-twist corrections at low $Q^2$ scales.

The low-$Q^2$ nucleon structure given in this work and the high-twist 
QCD corrections can be further studied and tested on 
the future high-energy facilities, such as 
EIC \cite{AbdulKhalek:2021gbh,Accardi:2012qut}, 
EicC \cite{Chen:2018wyz,Chen:2020ijn,Anderle:2021wcy,Wang:2022xad}, 
and JLab 24 GeV program \cite{Accardi:2023chb}. 
In order to access the small $Q^2$ data, the electrons 
of very small angles should be measured efficiently. 
Judged by the difficulty of detecting the small-angle electrons, 
the low-energy EicC is a little more appropriate  
to access the low-$Q^2$ data on the nucleon structure. 
More high-precision experiential data are looking forward 
to finally solve the problems regarding the transition 
from the nonperturbative region to the perturbative region.

%%\begin{acknowledgments}
%%\end{acknowledgments}

\bibliographystyle{apsrev4-1}
\bibliography{refs}

\end{document}